\documentclass[amsmath,10pt,nofootinbib,prl,twocolumn,superscriptaddress, bibliography]{revtex4-2}
\usepackage{siunitx} 
\usepackage{microtype} 
\usepackage{amsmath}
\usepackage{amsbsy}
\usepackage{amssymb}
\usepackage{graphicx}
\usepackage{color}
\usepackage{physics}
\usepackage{soul}
\usepackage{bm}
\usepackage{array}
\usepackage{multirow}
\usepackage{lipsum}
\usepackage{nameref}
\usepackage[capitalize]{cleveref}
\usepackage[normalem]{ulem}
\usepackage[normalem]{ulem} 
\usepackage[caption=false]{subfig}
\usepackage{xr}
\makeatletter
\newcommand*{\addFileDependency}[1]{
  \typeout{(#1)}
  \@addtofilelist{#1}
  \IfFileExists{#1}{}{\typeout{No file #1.}}
}
\makeatother

\newcommand*{\myexternaldocument}[1]{%
    \externaldocument{#1}%
    \addFileDependency{#1.tex}%
    \addFileDependency{#1.aux}%
}

\myexternaldocument{supplement}

\usepackage{verbatim}
\usepackage[bottom]{footmisc}

\makeatletter
\newcommand{\quickwordcount}[1]{%
  \immediate\write18{texcount -1 -sum -merge #1.tex > #1-words}%
  \immediate\openin\somefile=#1-words%
  \read\somefile to \@@localdummy%
  \immediate\closein\somefile%
  \setcounter{wordcounter}{\@@localdummy}%
  \@@localdummy%
}
\makeatother

\usepackage{natbib}

\begin{document}
\title{Elucidating the Role of Filament Turnover in Cortical Flow \\using Simulations and Representation Learning}
\author{Yuqing Qiu}
\affiliation{Department of Chemistry, \\University of Chicago, Chicago, IL 60637 USA}
\affiliation{James Franck Institute, University of Chicago, Chicago, IL 60637 USA}

\author{Elizabeth D.\ White}
\affiliation{Graduate Program in Biophysical Sciences, University of Chicago, Chicago, IL 60637 USA}

\author{Edwin M.\ Munro}
\affiliation{Graduate Program in Biophysical Sciences, University of Chicago, Chicago, IL 60637 USA}
\affiliation{Department of Molecular Genetics and Cell Biology, University of Chicago, Chicago, IL 60637, USA}

\author{Suriyanarayanan Vaikuntanathan$^*$}
\affiliation{Department of Chemistry, \\University of Chicago, Chicago, IL 60637 USA}
\affiliation{James Franck Institute, University of Chicago, Chicago, IL 60637 USA}
\affiliation{Graduate Program in Biophysical Sciences, University of Chicago, Chicago, IL 60637 USA}

\author{Aaron R.\ Dinner$^\dagger$}
\affiliation{Department of Chemistry, \\University of Chicago, Chicago, IL 60637 USA}
\affiliation{James Franck Institute, University of Chicago, Chicago, IL 60637 USA}
\affiliation{Graduate Program in Biophysical Sciences, University of Chicago, Chicago, IL 60637 USA}


\begin{abstract} 
Cell polarization relies on long-range cortical flows, which are driven by active stresses and resisted by the cytoskeletal network. 
While the general mechanisms that contribute to cortical flows are known, a quantitative understanding of the factors that tune flow speeds has remained lacking. 
Here, we combine physical simulation, representation learning, and theory to elucidate the role of actin turnover in cortical flows. 
We show how turnover tunes the actin density and filament curvature and use representation learning to demonstrate that these quantities are sufficient to predict cortical flow speeds.  We extend a recent theory for contractility to account for filament curvature in addition to the nonuniform distribution of crosslinkers along actin filaments due to turnover. We obtain formulas that can be used to fit data from simulations and microscopy experiments.
Our work provides insights into the mechanisms of contractility that contribute to cortical flows and how they can be controlled quantitatively.

\end{abstract}

\maketitle

\begingroup\renewcommand\thefootnote{$^*$}
\footnotetext{Co-corresponding author. Email: svaikunt$@$uchicago.edu}
\endgroup
\begingroup\renewcommand\thefootnote{$^\dagger$}
\footnotetext{Co-corresponding author. Email: dinner$@$uchicago.edu}
\endgroup

\section{Introduction}
The cortex is a protein network that is associated with the inner face of the plasma membrane of most animal cells \cite{bray1988cortical,chugh2018actin,kelkar2020mechanics}. 
Within the cortex, assemblies of non-muscle myosin II motors generate tension by pulling on actin filaments.
This tension determines cell shape and cell-cell interactions, and its gradients can give rise to long-range flows that transport proteins for cell polarization \cite{munro2004cortical,lang2017proteins}, migration \cite{hernandez2017polarized}, and division~\cite{carvalho2009structural}.  
These dynamics are regulated by the network architecture~\cite{mak2016interplay,ronceray2016fiber,Stam_Freedman_Banerjee_Weirich_Dinner_Gardel_2017,malik2019scaling} and actin filament assembly and disassembly (turnover) \cite{hiraiwa2016role, mcfadden2017filament,banerjee2017actomyosin, mccall2019cofilin, malik2019scaling,foster2022active}. 

From an active fluids perspective, the emergence of long-range cortical flow can be attributed to the buildup of localized active stress, which works against the viscous resistance of the network~\cite{malik2019scaling}. 
Experiments that reconstitute actin networks \textit{in vitro} \cite{mccall2019cofilin} and simulations \cite{mcfadden2017filament} showed that actin turnover tunes the viscosity by fluidizing the network. The contractility underlying the active stresses also depends on actin turnover, as well as actin density~\cite{banerjee2017actomyosin,malik2019scaling,foster2022active}.  
At the same time, it remains unclear whether actin filament buckling, which has been identified as a mechanism of contractility in mixtures of preformed filaments
\cite{liverpool2009mechanical,murrell2012f,lenz2012contractile,ronceray2016fiber,belmonte2017theory,Stam_Freedman_Banerjee_Weirich_Dinner_Gardel_2017}, is a major source of contractility in the presence of actin turnover. 
While we thus have a qualitative understanding of how network structure and dynamics affect cortical tension and, in turn, flow, a quantitative understanding remains lacking.

One major challenge in interpreting both simulations and experiments of cytoskeletal systems is identifying collective variables that capture the physics. 
Machine learning in principle holds promise for identifying such variables, and more generally for discovering physical models~\cite{carleo2019machine}. Recently, deep neural networks have been used to infer and predict the biophysical dynamics of actomyosin networks~\cite{liu2022cellular,qu2022three,colen2021machine,yoshida2023machine,schmitt2023zyxin,alderfer2022morphological}.  
Machine learning can be an ideal tool to assist in discovering and quantitatively characterizing mechanisms of contractility in different contexts. Nevertheless, connecting the information decoded by machine learning from simulations and experiments to physical theories remains challenging.

In this study, we use simulations, representation learning, and theory to investigate the factors that tune long-range cortical flow during anterior-posterior polarity establishment in the early {\it C.\ elegans} embryo~\cite{munro2004cortical}. 
We perform simulations that reproduce experimentally observed trends in cortical flow speed as actin turnover rates vary. 
Analyzing the simulations with representation learning and dimensional reduction techniques, we identify a latent representation of the contractile flow. We show how this latent representation is consistent with a microscopic model of contractility and motivates its extension. Our work thus provides insights into mechanisms of contractility underlying cortical flow and, more generally, shows how machine learning can be used to guide the development of physical models. 

\section{Simulations reproduce trends in cortical flow speeds as actin turnover rates vary}
We focus on exploring how the activity from filament turnover couples to a myosin gradient to control force generation for long-range flow.  
Previous experimental work revealed how specific actin-binding proteins, such as formin, cofilin, plastin, and profilin shape cortical flows~\cite{pollard2000molecular,bugyi2010control,kelkar2020mechanics}.
In particular, profilin blocks the assembly of actin filaments at their pointed ends~\cite{pollard1984quantitative} and promotes elongation at their barbed ends~\cite{goode2007mechanism}. 
Recently, some of us investigated how the level of profilin tunes the magnitude of the anterior directed flow through the control of actin treadmilling in {\it C.\ elegans} embryos~\cite{white2023uncovering}. These experiments showed that cortical flow speed depends nonmonotonically on profilin expression levels~\cite{white2023uncovering}. 

\begin{figure*}[tbp]
\centering
\includegraphics[width=\textwidth,clip=true]{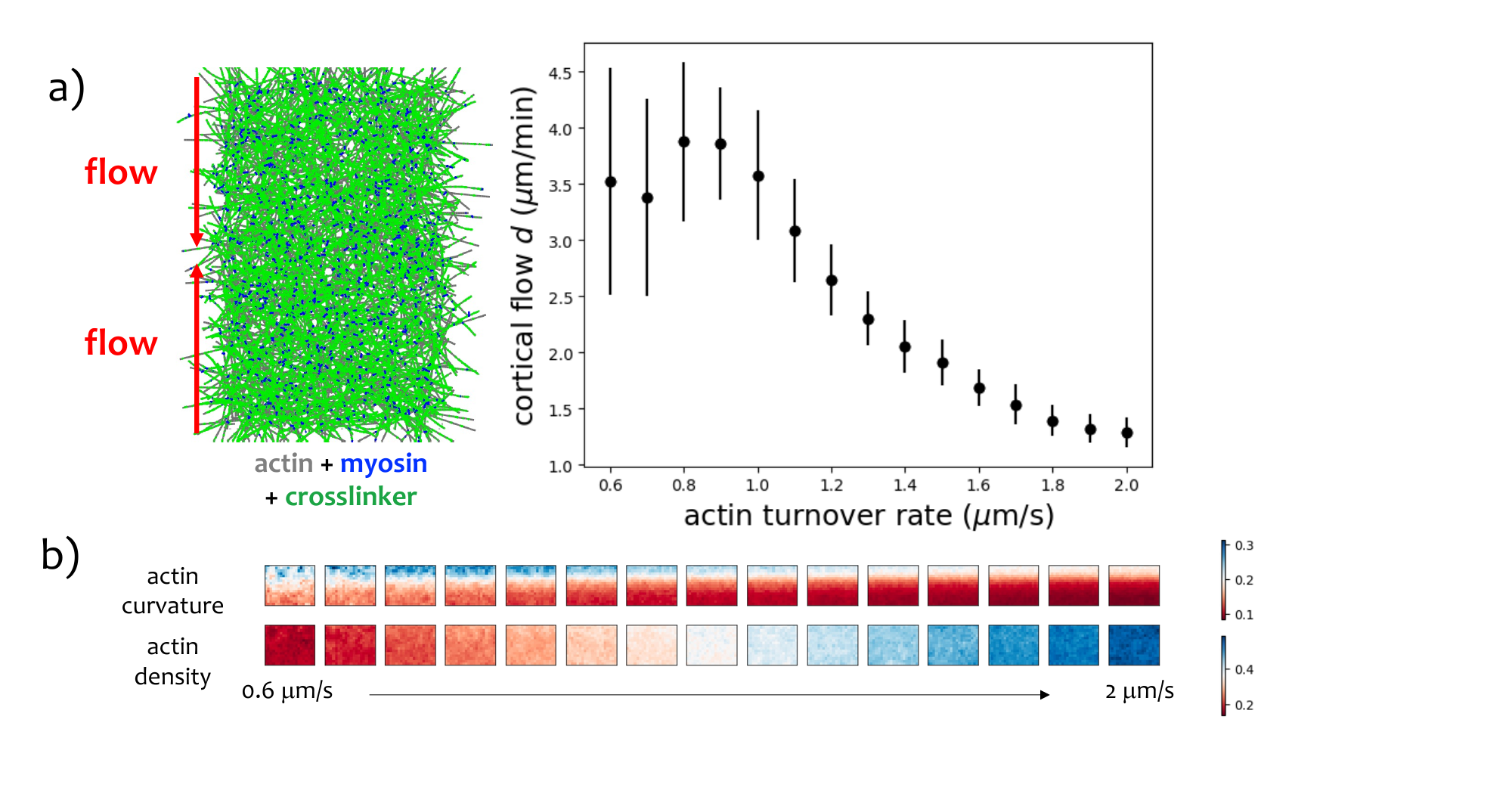}
\caption{Actin turnover modulates flow speed, actin density, and filament curvature. 
(a) 
Snapshot of simulation box and the cortical flow as a function of assembly/disassembly (turnover) rate. (b) Maps depicting the average filament curvature and density from simulations. These property maps are generated by averaging the data from the top and bottom halves of the simulation boxes in (a), and the upper part of the maps corresponds to the center of the simulation box.}
\label{fig:exp_vs_sim}
\end{figure*}

The methods that we employ below can be applied directly to experimental data. 
However, because the experimental data are limited and their interpretation is subject to assumptions, here we analyze data from agent-based simulations that are parameterized to reproduce the experimental trends quantitatively \cite{white2023uncovering}. The simulations are in two dimensions and include actin filaments, myosin motors, and actin crosslinking proteins~\cite{CytosimGit}. We model filament turnover by allowing filaments to shrink and grow with defined rates~\cite{white2023uncovering}.
Each nucleated filament grows at a fixed actin assembly rate $k$ for an average of 8.5 s before terminating growth, and then the actin filament shrinks from the pointed end with a disassembly rate equal to the assembly rate. Using this protocol, higher assembly/disassembly rates generate longer actin filaments, mimicking the effects of various levels of profilin in experiments.  
Previously, we found that setting the filament nucleation rate to 121 s$^{-1}$ and the actin assembly/disassembly rate to 1.5 $\mu$m/s reproduces the actin densities and average filament lifetimes of wild-type embryos~\cite{li2021filament,white2023uncovering}.
Given these rates, we tuned the number of crosslinkers to reproduce the inferred bundle size distribution and construct a gradient in myosin to drive cortical flow at experimentally measured speeds for wild-type embryos \cite{white2023uncovering}.
We then performed simulations for a range of actin assembly/disassembly rates and obtained flow speeds (\cref{fig:exp_vs_sim}a) that are in reasonable quantitative agreement with measured values in embryos in which profilin expression is reduced~\cite{white2023uncovering}.


\section{A latent representation reveals the physics of cortical flow}
Our goal is to develop a predictive physical model of cortical flow in terms of measurable quantities.  As noted above, cortical flows are thought to arise from a balance of active stress and network viscosity, both of which depend on actin density~\cite{malik2019scaling,banerjee2017actomyosin,foster2022active}.  An additional quantity that can be obtained from video microscopy is filament curvature, a proxy for filament buckling, a mechanism of contractility~\cite{lenz2012contractile,Stam_Freedman_Banerjee_Weirich_Dinner_Gardel_2017,murrell2012f,ronceray2016fiber,belmonte2017theory}.  
To determine if the measurable density and curvature are sufficient to predict cortical flow without presupposing a model, we use these quantities as inputs to a machine learning procedure.

\begin{figure*}[htbp]
\centering
\includegraphics[width=\textwidth,clip=true]{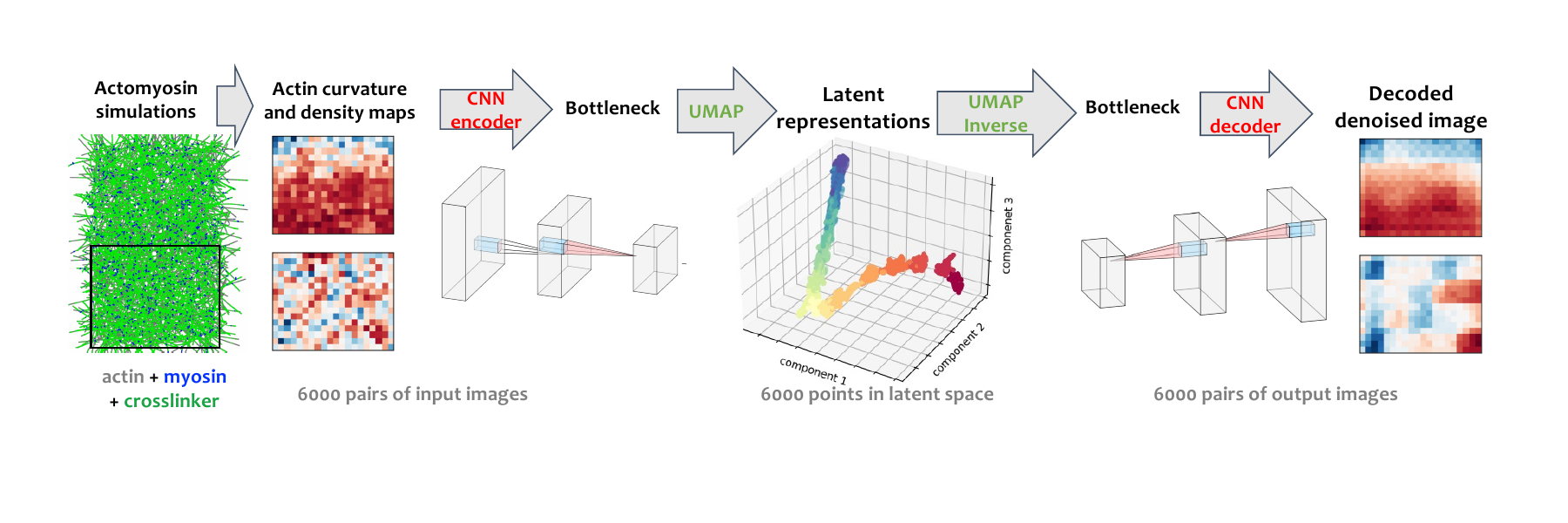}
\caption{Machine-learning workflow. The training data are composed of two channels: filament curvature and density maps. The input and output have a dimension of (2,16,20), while the CNN bottleneck has a dimension of (2,4,5). 6000 pairs of input images are used, generated from simulations at 15 actin turnover rates, saving 400 frames for each condition as described in SI Section S1A). Each point in the latent space shown corresponds to a pair of input images and is colored according to the simulation condition.}
\label{fig:workflow}
\end{figure*}

We grid the simulation region into 1 $\mu$m $\times$  1 $\mu$m boxes and compute the average actin density and filament curvature in each box (\cref{fig:workflow} and S1 Methods A).
As the assembly/disassembly rate increases, in general, the actin density increases and the curvature decreases (\cref{fig:exp_vs_sim}b).  The latter trend is a result of an increase in the average filament length, which tends to decrease the average number of motors and crosslinkers per unit length for fixed total numbers of motors and crosslinkers.

Motivated by recent work \cite{colen2021machine,schmitt2023zyxin}, we use a convolutional neural network (CNN) to analyze these data. Specifically, we apply the CNN as an autoencoder to extract features and denoise the data. 
The network architecture and training procedure are described in SI Section S1B.
In this approach, we train the network to reconstruct each 640-dimensional input (a stack of two images of filament curvature and density, with each image comprising 16 $\times$ 20 pixels) after passing it through a 40-dimensional bottleneck that defines a latent representation; to visualize this still many-dimensional latent representation,  
we use Uniform Manifold Approximation and Projection (UMAP)~\cite{leland2018umap} to further compress the information to three dimensions (3D). UMAP should preserve the global topological structure of the input~\cite{leland2018umap}. The procedure is summarized in \cref{fig:workflow}.  

\begin{figure}[tbp]
\centering
\includegraphics[width=0.48\textwidth,clip=true]{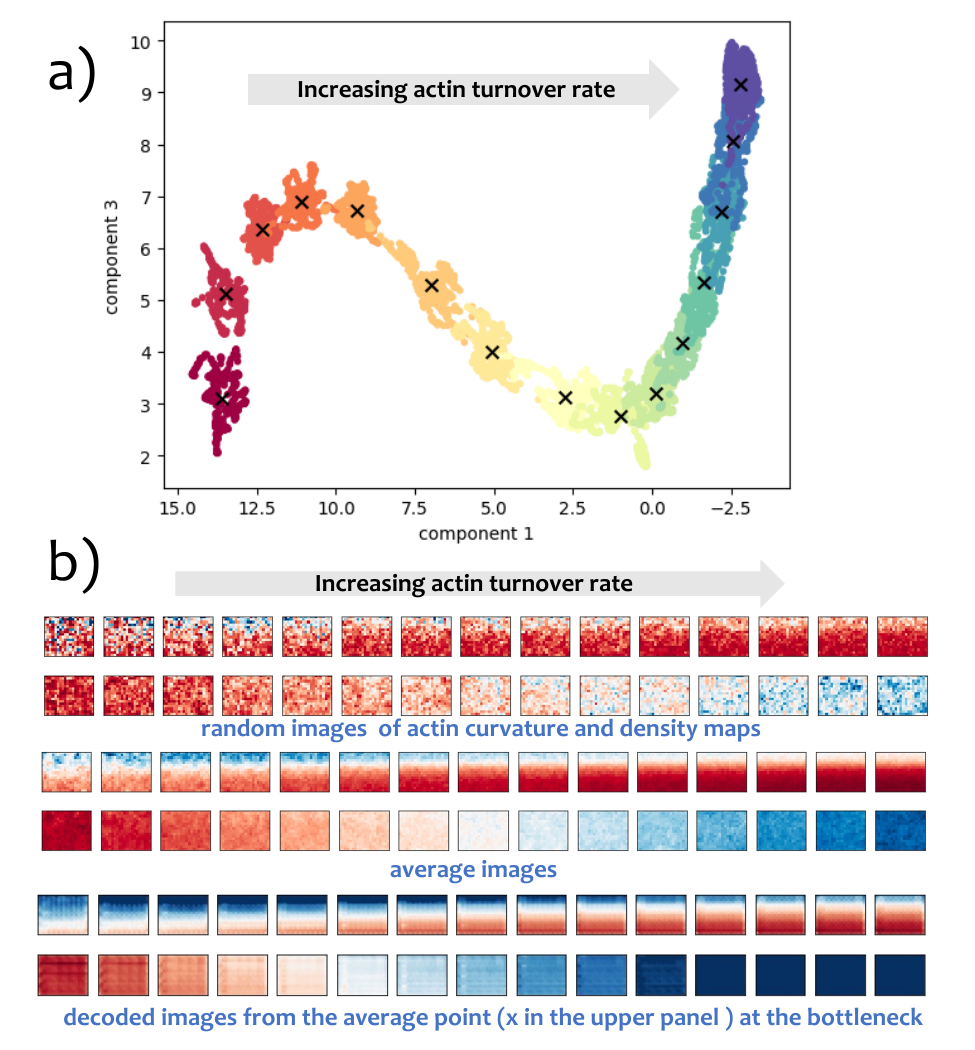}
\caption{A latent representation that reveals trends in actin density and filament curvature. a) 3D latent representation generated by the workflow in~\cref{fig:workflow}, colored by the assembly/disassembly rates in actin turnover ranging from 0.6 (red) to 2 (blue) $\mu$m/s. The 3D latent representation is projected to the 2D plane of components 1 and 3 for visualization. Black crosses are the average points for each actin turnover condition in the latent space. b) Comparison of random images, average images of 400 snapshots under each actin turnover condition, and the representative images decoded from the black crosses at the bottleneck in panel a.}
\label{fig:BN}
\end{figure}

We plot processed simulation data for each assembly/disassembly rate in a different color in \cref{fig:BN}.  The color clustering suggests that the procedure captures features that vary systematically with the assembly/disassembly rates in actin turnover. 
Similar clusters are seen when training with only one of the two features (SI Section S2C), consistent with the fact that both features vary with the assembly/disassembly rate of actin turnover. However, the same global topological structure in the latent space is not obtained (SI Section S2C), demonstrating that both actin density and curvature are needed to model flow. Therefore, we focus on interpreting the latent space from training using both actin density and curvature.
We find the average of each cluster in the latent space and construct the corresponding representative denoised actin density and curvature maps for further visualization and computation by training 
a fully connected neural network to invert the UMAP projection (SI Method S1B) (\cref{fig:BN}). Images decoded from the latent space (Figure 3B) more clearly reveal trends than simply averaging the input data from each condition (compare Fig.\ \ref{fig:decoded_data} with Fig.\ S2). 

A striking feature of the curve connecting the cluster averages in \cref{fig:BN} is that it has two extrema, near elongation rates 0.85 and 1.5 $\mu$m/s.  
To relate these features to physical quantities, we analyze the denoised images. These show that the first extremum is at the actin turnover rate where the filament curvature is maximized (\cref{fig:decoded_data}a). The second extremum corresponds to a fast assembly/disassembly rate in actin turnover that produces a high actin density and minimizes the relative variance in actin density. The fact that these assembly/disassembly rates are close to the value that reproduces the density and cortical flow speed of wild-type embryos \cite{li2021filament} suggests that there may be selection pressure to maintain network homogeneity. 
Our interpretation of the extrema aligns with the results of training with each feature by itself:  only the first extremum is obtained in the latent representation from training with filament curvature, while only the second extremum is obtained in the latent representation from training with actin density (SI Fig.\ S3). 

Besides the two extrema, one other feature that stands out in the latent representation is that the two clusters at the lowest assembly/disassembly rates are separate from the others. These assembly/disassembly rates produce high variances of actin density and unstable cortical flows (SI S2B), in agreement with experimental observations~\cite{white2023uncovering}.  The features identified by the machine-learning workflow guide the construction of a physical model of cortical flow, as we now describe.

\section{An active fluid model provides insights into the effects of actin turnover on cortical flow}\label{sec:theory}


We now show that the features that we demonstrated to be important for predicting flow immediately above, actin density and curvature, are consistent with a minimal physical model for contractility. 
Following previous work~\cite{banerjee2017actomyosin, malik2019scaling, foster2022active}, we define the active stress tensor $\mathbf {\Sigma_{a}} = \Sigma_{a} \mathcal{I}$ and the viscous stress tensor $\mathbf {\Sigma_{v}} = \eta {\nabla}\cdot u \mathcal{I}$, where $u$ is the velocity field and $\mathcal{I}$ is the identity tensor.  Balancing these stresses and using the divergence of the velocity field as a measure of contractility, we obtain
\begin{equation}
\label{eq:sigma_eta_ratio}
    d \equiv - {\nabla}\cdot u = \Sigma_{a}/\eta.
\end{equation}

To understand how actin turnover modulates contractility, we need to relate the magnitudes of the active stress $\Sigma_{a}$ and the viscosity $\eta$ to filament curvature and density. We do so by building on a recent model that accounts for the microscopic dynamics of filaments connected by motors and crosslinkers ~\cite{furthauer2019self,foster2022active} (SI Section S2E). 
In addition to the theory in previous work~\cite{furthauer2019self,foster2022active},we consider filament curvature and take into account the orientation of filament segments. This gives rise to an additional term in the expression for the active stress. 

We now summarize the theory. First, we show that 
\begin{equation}
\begin{split}
\label{eq:eta}
    \eta &\propto \rho^2 (C^M_0\gamma^M  + C^X_0\gamma^X),
\end{split}
\end{equation}
where $\rho$ denotes the actin density, $\gamma^{X/M}$ is the friction force coefficient for crosslinkers (X) and motors (M), $C^{X/M}_0$ is the average concentration of crosslinkers/motors bound to each actin filament.
As $ \rho C^{X/M}_0$ represents the total concentration of bound crosslinkers/motors in the system, \cref{eq:eta} physically derives from the fact that actin filaments (proportional to $\rho$) work against the friction from the crosslinkers and motors that connect them to surrounding filaments ($\rho C^{X/M}_0\gamma^{X/M}$). 

Then, we show that the active stress can be decomposed into a component that depends on buckling (i.e., curvature), $\Sigma^{b}_{a}$, and another that depends on actin turnover, $\Sigma^{t}_{a}$:
\begin{equation}
\label{eq:stress_active}
\Sigma_{a} = \Sigma^{b}_{a} + \Sigma^{t}_{a}
\end{equation} 
with 
\begin{equation}
\begin{split}
\label{eq:detailed}
    \Sigma^{b}_{a}  &\propto \frac{\lambda}{2}C_0^X \gamma^X  \rho ^2 V_{\|}  \frac{ C_0^M \gamma^M}{C_0^M \gamma^M + C_0^X \gamma^X},\\
    \Sigma^{t}_{a}  &\propto -C_1^X \gamma^X  \rho ^2 V_{\|}  \frac{ C_0^M \gamma^M}{C_0^M \gamma^M + C_0^X \gamma^X} .\\
\end{split}
\end{equation}
where $\lambda$ is a scalar that increases with curvature, and $V_{\|}$ is the magnitude of motor head velocity. 
Our previous work demonstrates that actin filament assembly rate can tune the composition of crosslinkers bound to actin~\cite{freedman2019mechanical,qiu2021strong}. 
Similarly, here, the simulations indicate that actin turnover modulates the distribution of crosslinkers/motors on actin filaments (SI Section S2F): assembly at the barbed end and disassembly at the pointed end biases crosslinkers toward the pointed end (i.e., older segments). 
$C_1^X$ is the deviation of the concentration from uniformly distributed, where its sign reports whether the accumulation of crosslinkers is at the barbed or the pointed end. 
$C_1^X$ is negative in our system due to the accumulation of crosslinkers toward the pointed ends, and $\Sigma^{t}_{a}$ contributes positively to the active stress. 

As explained in the SI, the configurations contributing to the stress $\Sigma^{t}_{a}$ can also contribute to buckling. Consider for example, two filaments in an
antiparallel configuration. The theoretical analysis outlined here (and detailed in the SI) shows that the translational velocity of an actin filament is proportional to the motor head velocity and is directed towards its pointed end. The translational velocity of actin filaments hence drives
the sliding of two filaments passing each other. The accumulation of crosslinkers at the pointed end can serve to hinder sliding, leading to buckling.

\begin{figure}[h!]
\centering
\includegraphics[width=0.46\textwidth,clip=true]{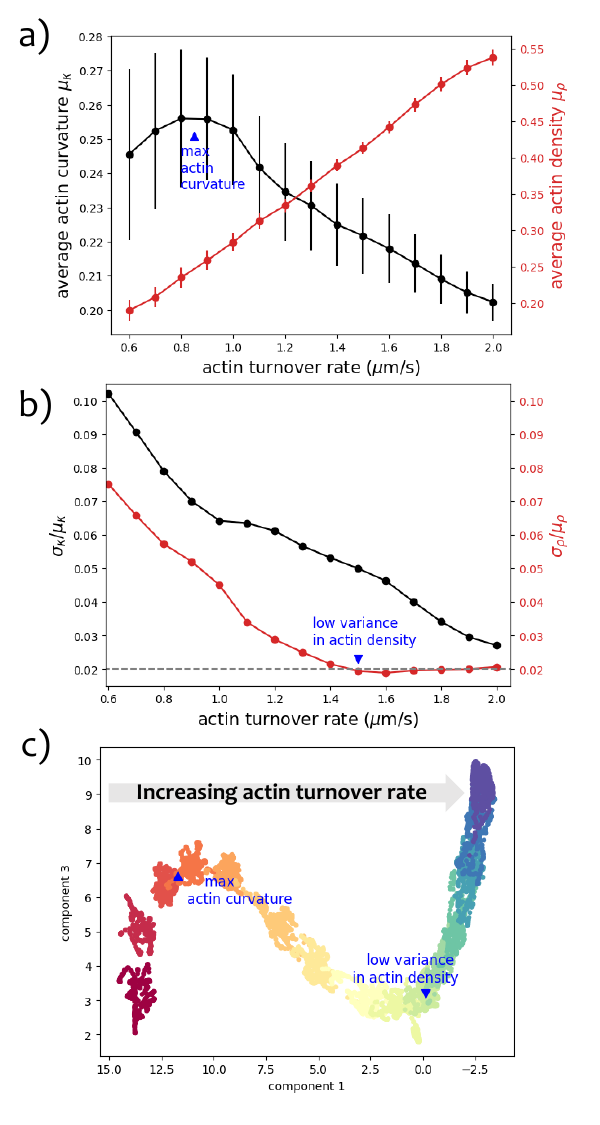}
\caption{Interpretation of the two extrema in the latent representation using the decoded images in \cref{fig:BN}b. a) Filament curvature $\mu_\kappa$ measured at the maximum of the myosin gradient (center of the simulation box) and average actin density $\mu_\rho$ of the entire simulation cell. Curves depict the means $\mu_\kappa$ and $\mu_\rho$ and error bars show the standard deviations $\sigma_\kappa$ and $\sigma_\rho$.  b) Relative variance of filament curvature $\sigma_\kappa/\mu_\kappa$ and actin density $\sigma_\rho/\mu_\rho$. c) The first extremum at actin turnover of 0.85 $\mu m/s$ corresponds to the maximum actin buckling (black curve in panel a). The second extremum at actin turnover of 1.5 $\mu m/s$ corresponds to the minimum speed with variance in actin density that is within 2\% of the magnitude of actin density (red curve in panel b).}
\label{fig:decoded_data}
\end{figure}

\begin{figure}[tbp]
\centering
\includegraphics[width=0.5\textwidth,clip=true]{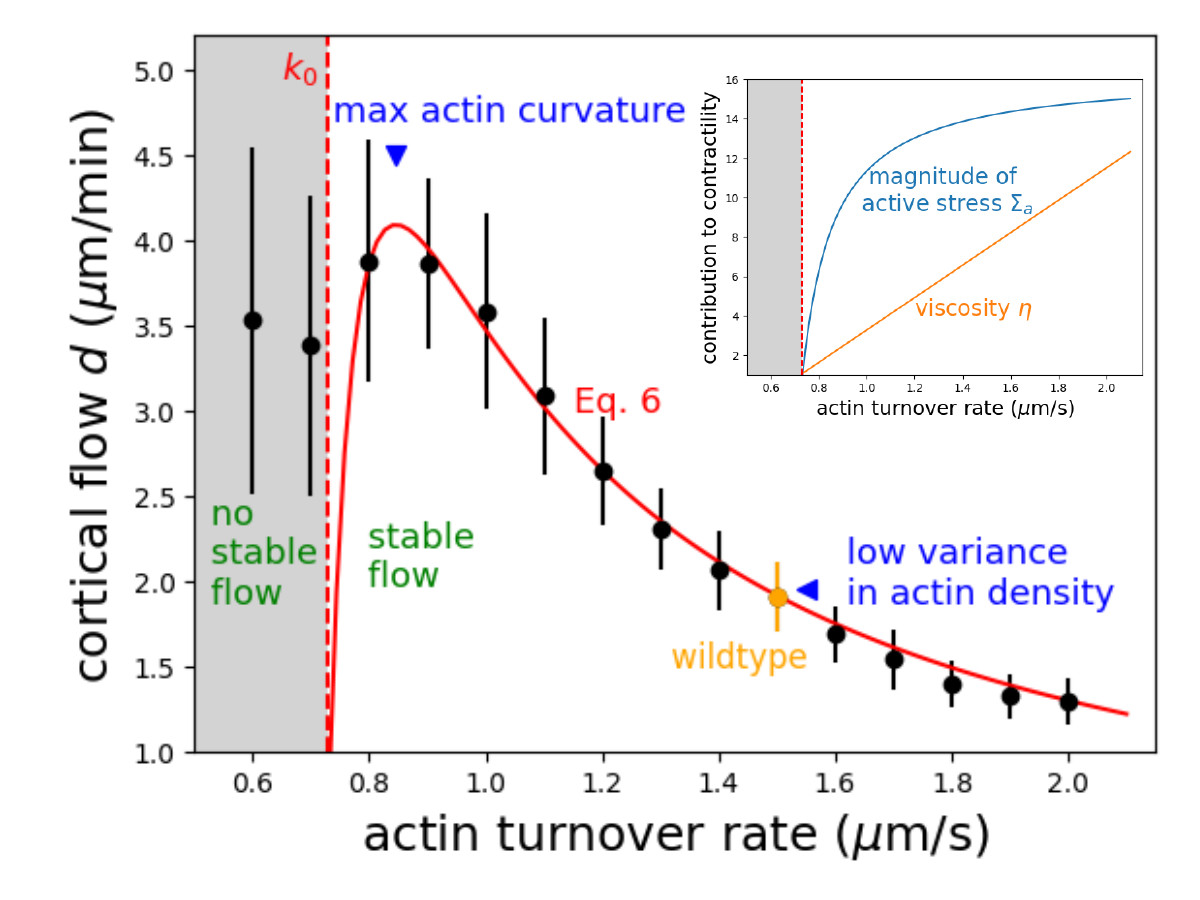}
\caption{ The physical model (red curve) is obtained by fitting \cref{eq:final} to stable flow rates generated by actin turnover above the threshold identified by the latent representations. The dashed red line marks the assembly/disassembly rate threshold $k_0$ obtained in the fitting. The gray box denotes the actin turnover conditions that produce unstable flow. (inset) The contributions to contractility from active stress (blue) and viscosity (orange) at the parameters estimated from the fit. }
\label{fig:fit_only}
\end{figure}

Substituting these results into \cref{eq:sigma_eta_ratio}, the contractility can be written as
\begin{equation}
\label{eq:d}
d \propto \gamma^XV_{\|} \frac{ (\frac{\lambda}{2}C_0^X  - C_1^X)C^M_0 \gamma^M}{(C^M_0 \gamma^M + C_0^X \gamma^X)^2}.
\end{equation}
Since the overall concentration of myosin motors $\rho_{M}$ is a constant in our simulations, the concentration of myosin motors along filaments (i.e., per unit length) is inversely proportional to the density of filaments $\rho$.   
Inspired by the latent representation in \cref{fig:decoded_data}c, we include a threshold $\rho_0$ to account for the limit below which the network loses connectivity and express $C^M_0 = \rho_{M}/(\rho-\rho_0)$.
We showed in~\cref{fig:decoded_data}a that the actin density is essentially proportional to the assembly/disassembly speed $k$ in actin turnover, $\rho = \epsilon k $. Inserting these expressions into \cref{eq:d}, we can write the flow rate in~\cref{eq:sigma_eta_ratio} as
\begin{equation}
\begin{split}
\label{eq:final}
    d  &= \frac{ \alpha (k -k_{0})}{[1 + \beta (k -k_{0})]^2}, \\
    \alpha &= \frac{V_{\|} \gamma^X \epsilon }{\gamma^M \rho_{M}} \left(\frac{\lambda}{2}C^X_0  - C^X_1\right),\\
    \beta &= \frac{ \gamma^X \epsilon C^X_0 }{\gamma^M \rho_{M}}.
\end{split}
\end{equation}
$\alpha$, $\beta$ and $k_0$ are independent fitting parameters.
Our simulations, together with the latent representation, show that assembly/disassembly rates below 0.8 $\mu$m/s in actin turnover induce irregular flow (Fig.\ S4). Since the theoretical framework in \cref{eq:final} accounts for forces exerted by motors on a fully connected network, we fit it to the steady flows for assembly/disassembly rates above the 0.8 $\mu$m/s threshold. 
In this range of assembly/disassembly rates, the model describes the dependence of flow $d$ on the actin turnover perfectly (\cref{fig:fit_only}). The fitting indicates that the contribution to contractility from active stress initially increases and then plateaus at high assembly/disassembly rates in actin turnover (inset of \cref{fig:fit_only}). In contrast, viscosity contribution exhibits a linear increase. The interplay between these two contributions gives rise to the non-monotonic trend.

The maximum contractility 
\begin{equation}
\label{eq:dmax}
    d^{max} = \frac{ V_{\|} }{4}  \left(\frac{\lambda}{2}  - C^X_1/C^X_0\right)
\end{equation}
occurs at an assembly/disassembly rate of 
\begin{equation}
    k^* = k_0 + (\gamma^M \rho_{M})/( \epsilon \gamma^X  C^X_0 ).
\end{equation} 
We thus see that the maximum flow rate is controlled by a difference between $\lambda/2$, which reflects the curvature, and $C^X_1/C^X_0$, which reflects the non-uniform distribution of crosslinkers along the filaments and also enhances the buckling of actins when $C^X_1 < 0$ (see discussion above). 
In the theory, $C_1^X/C_0^X$ can vary in the range [--1/$L_a$, 1/$L_a$], where $L_a$ is the length of the actin filament, with the sign depending on whether the crosslinkers are in excess at the pointed or barbed ends. 
In our simulations, the maximized length $L_a$ that actin filaments can reach before disassembly falls within the range of 5 to 17 $\mu m$ (SI Section S2F).
In contrast to this, actin buckling always contributes positively, with values within the range of [0, 3/$L$], 
where $L$ is a filament segment length over which actin curvature is induced and we set $L = $ 0.2 $\mu$m in simulations. We estimate from our simulation that $\lambda/2 \approx  1/(10 L)$ (SI Section S2E).
This suggests that both terms are of similar magnitude and to engineer a higher flow rate, one can enhance the degree of actin buckling and/or promote the accumulation of crosslinkers at the pointed end (for $C^X_1 < 0$).

\section{Conclusions}
A competition between active stresses and internal resistance to deformations controls long-range cortical flows that polarize animal cells.  
In this work, we combined physical simulation, representation learning, and theory to understand how contributing microscopic mechanisms tune flow speeds quantitatively.   
The inputs to the representation learning are spatially resolved maps of the actin density and filament curvature at different assembly/disassembly rates; the representation learning demonstrates that these data are sufficient to predict flow speed and identifies key values of the assembly/disassembly rate that we associate with maximum average curvature and minimum density variance. The former corresponds to the fastest cortical flow. Our machine-learning approach infers that the contractility has a parametric form that depends on the actin density and buckling which are both modulated by actin turnover. 

The representation learning guides the extension of a minimal physical model for contractility. The model relates active and viscous stresses to molecular concentrations by explicitly integrating the forces for elemental geometries. Our model and simulations confirm that actin buckling is a dominant mechanism for generating active stress in highly interconnected actomyosin networks with low filament rigidity. 
Although the relationship between cortical flow and microscopic parameters of the cortex is inherently complex, our findings demonstrate that the coupling between actin density, actin buckling, motor activity, and the non-uniform distribution of crosslinkers along actin filaments recapitulates the effect of actin filament turnover on the rate of cortical flow. The methods developed in this work can be generalized and combined with either experimental imaging or computer simulations to investigate the physics of dynamic contractile networks.  

\acknowledgements{This work was supported by the National Science Foundation through awards MCB-2201235 and PHY-2317138 (the Center for Living Systems at the University of Chicago).  S.V. and Y.Q. were supported by the National Institute of General Medical Sciences of the NIH under Award No. R35GM147400.}


\bibliographystyle{unsrt}
\bibliography{references}

\end{document}


\begin{center}
\textbf{\Large Supporting Information for\\ ``Elucidating the Role of Filament Turnover in Cortical Flow \\using Simulations and Representation Learning"}
\end{center}

\setcounter{figure}{0}
\setcounter{table}{0}
\setcounter{equation}{0}
\setcounter{page}{1}
\setcounter{section}{0}

\renewcommand{\theequation}{S\arabic{equation}} 
\renewcommand{\thepage}{S\arabic{page}} 
\renewcommand{\thesection}{S\arabic{section}}  
\renewcommand{\thetable}{S\arabic{table}}  
\renewcommand{\thefigure}{S\arabic{figure}}

\section{Methods}

\subsection{Implementation of actin turnover in the Cytosim simulations package }

The simulations are constructed to reproduce the flow along the anterior-posterior (AP) axis of \textit{C.\ elegans} embryos during maintenance phase \cite{munro2004cortical}. 
This flow is driven by a myosin gradient and actin turnover, and we follow Ref.~\onlinecite{white2023uncovering} in how we generate these features in Cytosim simulations~\cite{Cytosim,CytosimGit} and in turn the overall simulation setup.
Simulations are performed in a 20 $\mu$m $\times$ 32 $\mu$m  rectangular box with periodic boundary conditions in both dimensions.  We initialize the simulations with 44,500 crosslinkers and 1,600 myosin motors.  Crosslinkers are recycled at a rate of s$^{-1}$ uniformly throughout the simulation box.  Motors are added throughout the entire simulation space at the rate of 30 s$^{-1}$ and at an additional rate of 180 s$^{-1}$ in the center 10 $\mu$m of the system, and they are removed uniformly at the rate of 210 s$^{-1}$. 
%
%
%
Actin filaments are nucleated at a rate of 121 s$^{-1}$, 
grow at a fixed rate $k$ for 8.5 s before terminating growth and shrink from the pointed end with rate $k$. We refer to $k$ as the assembly/disassembly rate in the main text. 

\begin{figure*}[hbp]
\centering
\includegraphics[width=0.65\textwidth,clip=true]{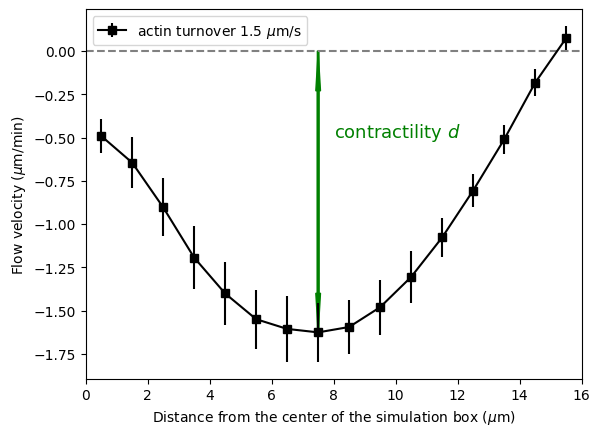}
\caption{Flow velocity along the X-axis of the simulation box. The black curve depicts the flow velocity profile. The contractility $d$ is defined as the magnitude of the fastest flow rate.}
\label{fig:flow_d}
\end{figure*}

This setup produces flow from the edges toward the center of the simulation box. Crosslinkers that are bound to the same actin filament in two consecutive frames are used as markers of the velocity. We average the velocity within 1 $\mu$m bins along the AP-axis, and we average corresponding bins from the two halves of the simulation box, reflected over the center. 
The contractility is the divergence of the velocity field, which here is simply the slope of the velocity profile.  To a first approximation, this slope is proportional to the maximum velocity (in the negative direction, \cref{fig:flow_d}), and we use the maximum velocity as a proxy for the contractility.

We compute the actin density and filament curvature in 1 $\mu$m $\times$ 1 $\mu$m bins and again average the two halves of the simulation box reflected over the center. Each actin filament is divided into 0.2 $\mu$m long segments. The actin density $\rho$ is computed as the number density of the filament segments. The curvature of each filament segment $\kappa$ is computed as the inverse of the radius of the circle that passes through the locations of the center segment and its two neighboring segments. The curvatures of the first and last segments in each filament are neglected. We average the curvatures in each bin. 

For each actin turnover rate $k$ ranging from 0.6 to 2.0 $\mu$m s$^{-1}$, we run 10 simulations of 90 s and save 1 frame per second. Only the last 40 frames of each simulation are used for all analyses. We average the flow rate over every 10 frames and thus collect 40 data points for flow rates at each actin turnover rate. 
We do not average the actin density and filament curvature maps for all frames and thus collect 400 data points for each.

\subsection{Architecture of Convolutional Autoencoder}
We implement the convolutional autoencoder in TensorFlow. The encoder is composed of two convolutional layers with ReLU activation functions, each of which has two filters of size 3 $\times$ 3 and is followed by a 2 $\times$ 2 max-pooling layer. 
The input data of the model has two channels corresponding to the actin density and curvature maps. Each map is a 16 $\times$ 20 image. The model is trained on a data set of 6000 frames from 15 actin turnover conditions. The pixel values of each channel are scaled by their maximum across all 6000 images so that the inputs are in the range [0, 1]. The maximum intensity of actin curvature is 4.79 $\mu$m$^{-1}$, and the maximum actin density is 176.5 $\mu$m$^{-2}$.  
The decoder is the same as the encoder but in reverse. It also contains two convolutional layers, followed by a final convolution layer with sigmoid activation functions that produce outputs in the range [0, 1]. The decoder accepts inputs with size (2,4,5) and uses convolutional layers to up-sample back to 16 $\times$ 20. 
The network is trained for 500 epochs and we use a 90 to 10 training-validation split on the input data set.

Once the model is trained, we freeze the weights in the model and insert UMAP~\cite{leland2018umap} in between the encoder and decoder to generate a 3D representation. We use min$\_$dist=0.075 and n$\_$neighbors=20 in the UMAP projection. Finally, fully connected dense layers are trained to map the 3D bottleneck back to the 40-dimensional input used for the decoder. This neural network is composed of 3 layers with tanh activation functions, containing 10, 20, and 40 nodes, respectively. In addition, a dropout layer with a dropout probability of 0.2 is added before the last dense layer. This network is trained for 500 epochs with a 0.1 validation split ratio. 
Combining the CNN autoencoder, UMAP, and the UMAP inverse neural network, we obtain the 3D latent representation of the actin feature maps. 

\section{Other Supporting Information}
\subsection{Average actin density and curvature from simulations}
\begin{figure}[hbp]
\centering
\includegraphics[width=0.65\textwidth,clip=true]{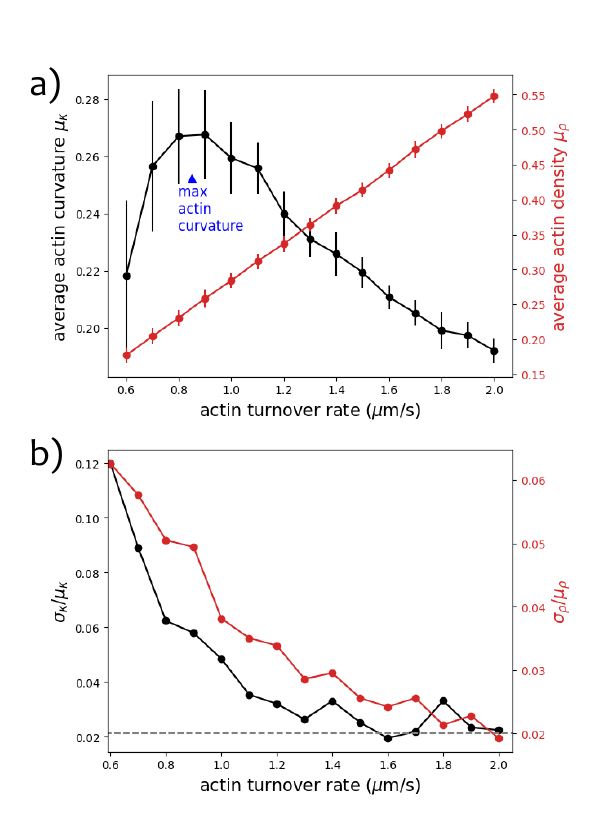}
\caption{Filament curvature and actin density  (Fig.\ 1b in the main text). a) Actin curvature $\mu_\kappa$ measured at the maximum of the myosin gradient (a 2 $\mu$m region at the center of the simulation box) and average actin density $\mu_\rho$ of the entire simulation cell. Error bars show standard deviations, $\sigma_\kappa$ and $\sigma_\rho$.  b) Relative variance of filament curvature $\sigma_\kappa/\mu_\kappa$ and actin density $\sigma_\rho/\mu_\rho$. }
\label{fig:compareF4}
\end{figure}

\cref{fig:compareF4} is the same as Fig.\ 4 of the main text, except that the analysis is performed on average images rather than denoised ones obtained from the computational pipeline. In general, the curves are not as smooth as in Fig.\ 4, and there is no clear feature near $k=1.5$ $\mu$m/s, which corresponds to the second extremum in the latent representation.


\subsection{3D bottleneck obtained from partial data } 
\begin{figure*}[hbp]
\centering
\includegraphics[width=1\textwidth,clip=true]{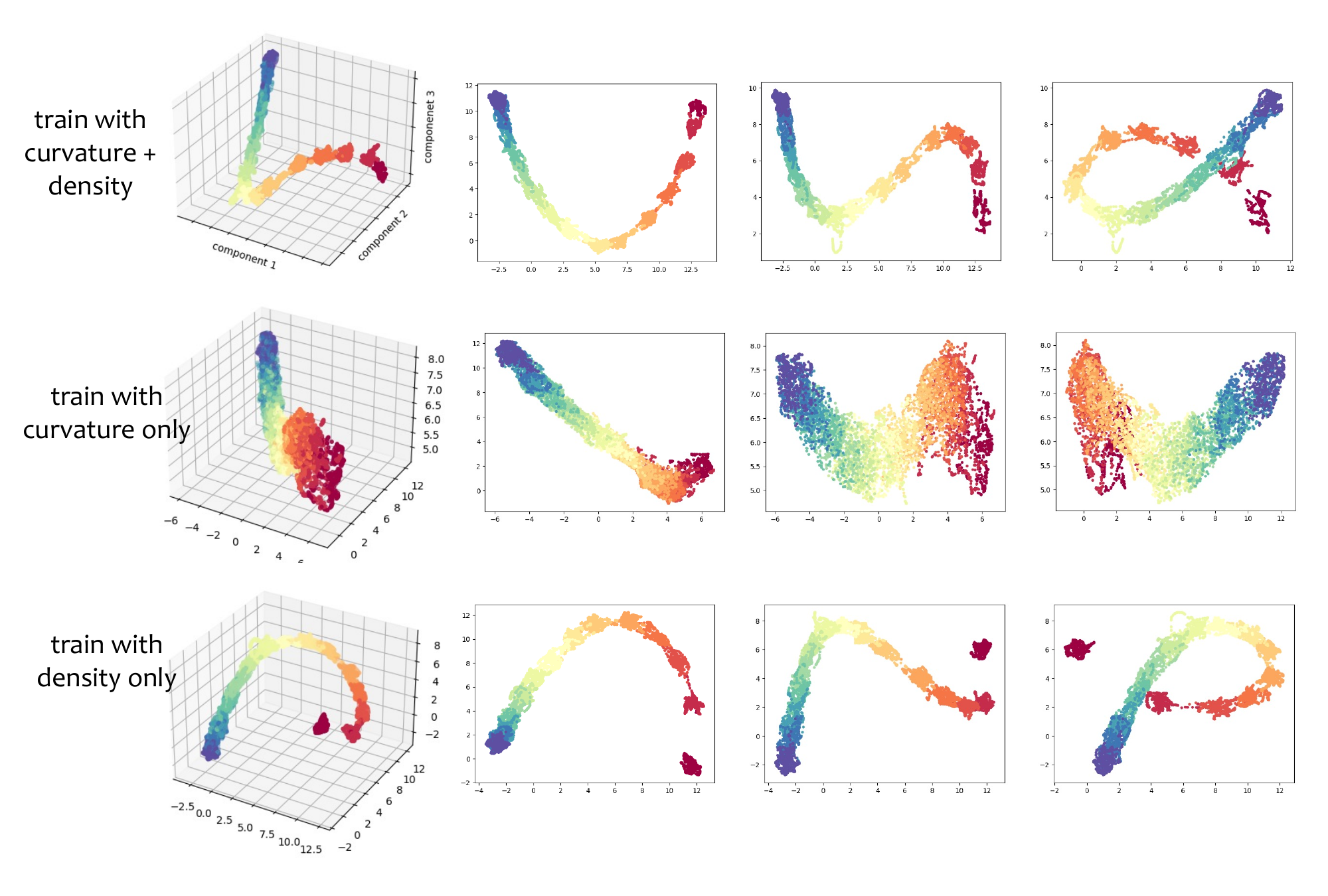}
\caption{3D bottleneck representations obtained from training with various data sets. The 3D bottleneck, generated through UMAP, is color-coded according to the simulation conditions, consistent with the main text. Each 3D bottleneck is followed by its corresponding projections onto the (component 1, component 2), (component 1, component 3), and (component 2, component 3) planes.}
\label{fig:training}
\end{figure*}

Here we compare the latent representation obtained from training solely with actin density or solely with filament curvature with that obtained from training with both.  \cref{fig:training} shows that the results depend on the data set. Notably, training with only actin curvature yields the first (in $k$) extremum and less separation between clusters representing different simulation conditions.
Training with only actin density leads to the appearance of the second extremum. 
This aligns with the interpretation in Fig.\ 4 of the main text, where the first extremum is assigned to the maximized actin curvature, and the second is interpreted as a consequence of the low variance in actin density.

\subsection{Cortical flow profiles at various actin turnover rates} 
\begin{figure*}[hbp]
\centering
\includegraphics[width=0.6\textwidth,clip=true]{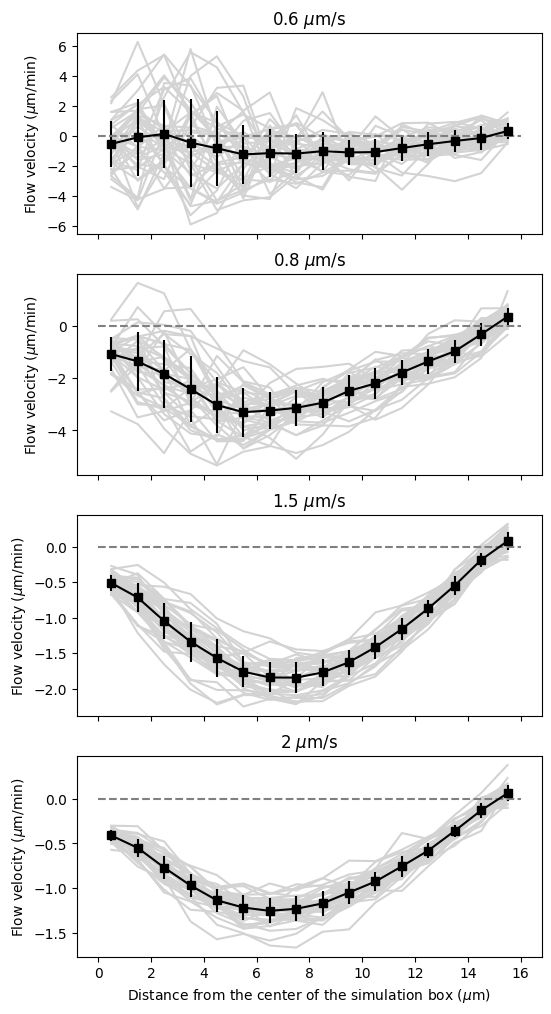}
\caption{Flow velocity profiles for the indicated assembly/disassembly rates. Gray curves are the flow velocity profile measured over the course of each 10 s interval. The black curves are the average flow velocity profiles. Error bars are standard deviations over 40 data points.  }
\label{fig:flow_detail}
\end{figure*}

\cref{fig:flow_detail} shows that the flow is unstable for low assembly/disassembly rates.
\clearpage

\subsection{Derivation of Eq.\ 1}
To connect the simulation flow rates to a theoretical description of the mechanism of contractility, we invoke force balance to write 
\begin{equation}
\label{eq:force balance}
\mathbf{\nabla} \cdot \mathbf{\Sigma} =   
  \mathbf{\nabla} \cdot (\mathbf {\Sigma_{a}} +    
  \mathbf{\Sigma_{v}} + \mathbf{\Sigma_{e})} = f_{D},
\end{equation}
where $\mathbf{\Sigma}$ is the total stress acting on the network, $\mathbf {\Sigma_{a}},  \mathbf {\Sigma_{v}}, \mathbf {\Sigma_{e}}$ are the active, viscous, and elastic network stresses, respectively, and $f_D$ is the friction from the surrounding fluid.
We can immediately simplify this expression because the fluid friction force $f_D$ is at least an order of magnitude lower than the contractile force generated by myosin~\cite{malik2019scaling}, allowing us to neglect hydrodynamics in the simulations. 
Also, the timescale of elastic stress has been shown to be 0.1~s \cite{malik2019scaling, white2023uncovering}, which is much shorter than the characteristic timescale of the cortical flow, $\sim$1 min; as a result, the network can be modeled as a viscous fluid, and $\mathbf {\Sigma_{e}}$ can be neglected. 
Thus, \cref{eq:force balance} reduces to  $\mathbf{\nabla} \cdot (\mathbf {\Sigma_{a}} + \mathbf {\Sigma_{v}}) \approx 0 $. 
The divergence of stress in a viscous fluid can be written as 
\begin{equation}
\label{eq:stress_1}
\mathbf{\nabla} \cdot \mathbf{\Sigma} = \mathbf{\nabla} \cdot (\mathbf \Sigma_{a}  +    \mathbf\Sigma_{v}) = \mathbf{\nabla} \cdot (\Sigma_{a} \mathcal{I}  +    \mathbf\Sigma_{v})  = 0 
\end{equation}
where $\Sigma_{a}$ is the magnitude of active stress. The velocity gradient in our simulations is only along the myosin gradient. The shear stress vanishes and there are only normal stresses along the direction of the flow.  Thus, $\mathbf\Sigma_{v}$ is a diagonal tensor. 
Using the divergence theorem,   
\begin{equation}
\label{eq:stress_2}
\int_V \mathbf{\nabla} \cdot \mathbf{\Sigma} dV= \int_V \mathbf{\nabla} \cdot (\Sigma_{a} \mathcal{I} +  \mathbf\Sigma_{v}) d V  = \int_{S} ( \Sigma_{a}\mathcal{I}  +  \mathbf\Sigma_{v}) \hat n dS = 0,
\end{equation}
where $S$ is the surface on which the stress acts on and $\hat n$ is its surface normal. 
One can write the viscous stress as the product of the network viscosity and the divergence of the velocity field $u$, $\mathbf\Sigma_{v} = \eta Tr({\nabla}\cdot u)\mathcal{I}$. Substituting, 
\begin{equation}
\label{eq:stress_2}  
\int_{S} (\Sigma_{a} \mathcal{I} +  \eta Tr ({\nabla} \cdot u)\mathcal{I}) \hat n dS = 0.
\end{equation}
This leads to $\Sigma_{a}  = -\eta Tr({\nabla}\cdot u)$ since the above equation holds for all $S$. We define the network contractility $d$ as the negative divergence of the velocity field and obtain
\begin{equation}
\label{eq:sigma_eta_ratio}
    d \equiv - Tr({\nabla}\cdot u) \equiv - \frac{\partial u}{\partial \hat n} = \frac{\Sigma_{a}}{\eta}.
\end{equation}

\subsection{A microscopic model for the active stress and viscosity }

We follow Ref.~\onlinecite{foster2022active} to derive the expressions of active stress $\Sigma_a$ and viscosity $\eta$ in Eq.\ 1 in the main text. We consider a system of actin filament segments of length $L$ that are connected by crosslinkers and motors. We measure the curvature of filaments in simulations from segments of 0.2 $\mu$m in length, as described above. We thus set $L =0.2$ $\mu$m. The total force exerted on filament $i$ through crosslinkers/motors (denoted by superscript $X/M$) is
\begin{equation}
F_i = \sum_j F_{ij}^{X/M}, 
\end{equation}
where $F_{ij}^{X/M}$ is the force exerted on filament segment $i$ by a crosslinkers/motors connected to filament segment $j$. 
We assume that the filament network is overdamped and that the motion of liquid surrounding the filament network can be ignored. The total force on filament segment $i$ is $F_{i} = 0$. 
We can also write the force density at position $\mathbf x$ as,
\begin{equation}
F(\mathbf x) 
= \sum_{ij} \delta (\mathbf x - \mathbf x_i) F_{ij}^{X/M}.
\end{equation}
As pointed out by Ref.~\cite{furthauer2019self}, in the absence of external force, $F(\mathbf x) = 0$. We define the material stress tensor as 
\begin{equation}\mathbf{\Sigma(x)} =  -\sum_{ij} \frac{(\mathbf x_{i}-\mathbf x_{j})}{2} F_{ij}^{X/M}  \delta (\mathbf x - \mathbf x_i),
\end{equation}
where $\mathbf x_{i}-\mathbf x_{j}$ denotes the distance between two connected filament segments and is assumed to be small. We can relate the force density to $\mathbf {\Sigma(x)}$ as follows.

\begin{equation}
\begin{split}
\label{eq:stress_force}
F(\mathbf x)& = \sum_{ij} \delta (\mathbf x - \mathbf x_i) F_{ij}^{X/M}\\ &= \frac{1}{2} \sum_{ij} \left[\delta (\mathbf  x - \mathbf x_i) F_{ij}^{X/M} + \delta (\mathbf x - \mathbf x_j) F_{ji}^{X/M}\right]\\ 
&= \frac{1}{2} \sum_{ij} \left[\delta (\mathbf  x - \mathbf x_i) F_{ij}^{X/M} - \delta (\mathbf x - \mathbf x_i + \mathbf x_i-\mathbf x_j) F_{ij}^{X/M}\right]\\
& = -\frac{1}{2}\sum_{ij}  F_{ij}^{X/M} \nabla \delta (\mathbf x - \mathbf x_i)(\mathbf x_{i}-\mathbf x_{j}) + \mathcal{O}\left((\mathbf x_{i}-\mathbf x_{j})^2\right) \\
&\approx \div \left[ -\sum_{ij} \frac{(\mathbf x_{i}-\mathbf x_{j})}{2} F_{ij}^{X/M}  \delta (\mathbf x - \mathbf x_i) \right] \\
&= \div   \mathbf {\Sigma}(\mathbf x).
\end{split}
\end{equation}
Note that the second equal sign holds because $F_{ij}^{X/M} = - F_{ji}^{X/M}$, and the fourth equal sign results from Taylor expansion of the $\delta$ function. 
Now we can integrate the force density $F(\mathbf x)$ to obtain expressions for the stress tensor $ \mathbf\Sigma(\mathbf x)$ and the active stress $\mathbf{\Sigma_a}(\mathbf x)$. 

Filament segment $i$ is characterized by the its center-of-mass position $\mathbf {x_i}$ and its local orientation $\mathbf {p}_i$. 
The distance between the binding position of a crosslinker/motor and the filament center is $s_i$. We denote the local orientation of filament $i$ at its center of mass as $\mathbf{p}^0_i$. 
The local orientation can be written as $\mathbf {p}_i(s_i) = \mathbf{p}^0_i + \int_0^{s_i} \frac{\partial \mathbf {p}_i }{\partial s_i}  d s_i$, where 
$\frac{\partial \mathbf {p}_i }{\partial s_i}$ marks the local curvature of the filament. We refer to this curvature term as $ \bm \kappa_i = \frac{\partial \mathbf {p}_i }{\partial s_i}$ and the local orientation can be expressed as $\mathbf {p}_i(s_i) = \mathbf{p}^0_i + \bm \kappa_i s_i$. Then, the binding position of a crosslinker is 
$\mathbf x_i + \int_0^{s_i} \mathbf p_i  ds_i = \mathbf x_i + s_i \mathbf p^0_i + \frac{\bm \kappa_i}{2} s_i^2$. 
The stress tensor at position $\mathbf x$ then can be written as:
\begin{equation}
\begin{split}
\label{eq:stress_integration}
\mathbf\Sigma(\mathbf x) = -\frac{1}{2} \sum_{ij}& {\int_{-\frac{L}{2}}^{\frac{L}{2}} \,ds_i} 
{\int_{-\frac{L}{2}}^{\frac{L}{2}} \,ds_j}  
{\int_{\Omega(\mathbf x)} \,d^3\mathbf x^{\prime}} \delta(\mathbf x_{i}-\mathbf x) \delta (\mathbf x_{j}-\mathbf x^{\prime})\times\\&\left(\mathbf x_i + s_i \mathbf p^0_i + \frac{\bm \kappa_i}{2} s_i^2 -\mathbf x_j - s_j \mathbf p^0_j  - \frac{\bm {\kappa}_j}{2} s_j^2 \right) F_{ij}^{X/M}  C^{X/M}(s_i, s_j),
\end{split}
\end{equation} 
where $\Omega(\mathbf x)$ is a sphere centered at $\mathbf x$, and its radius is the size of a crosslinker/motor,
$ C^{X/M}(s_i, s_j) $ is the concentration of crosslinkers/motors along the filaments (\textit{i.e.}, the number of crosslinkers/motors per unit length). We integrate all possible configurations of crosslinkers/motors connected to filament $i$ to obtain $\mathbf\Sigma(\mathbf x)$.
In what follows, we expand the expressions for $F^{X/M}_{ij}$ and $C^{X/M}(s_i, s_j)$ and then integrate to obtain $\mathbf\Sigma(\mathbf x)$.

The force $F^{X/M}_{ij}$ depends on the velocity difference between two crosslinker/motor heads $\Delta \mathbf v_{ij}^{X/M}$, and we use a friction coefficient $\gamma$ to relate force to velocity: 
\begin{equation}
\label{eq:force_velocity}
F^{X/M}_{ij} = -\gamma^{X/M} \Delta \mathbf v_{ij}^{X/M}.
\end{equation} 
The velocity difference $\Delta \mathbf v^{X/M}_{ij}$ can be written as:
\begin{equation}
\begin{split}
\label{eq:vij}
\Delta \mathbf v^X_{ij} &=  (\mathbf{v}_i + s_i\mathbf{\dot p}_i) - (\mathbf{v}_j + s_j\mathbf{\dot p}_j)=  (\mathbf{v}_i + s_i\mathbf{\dot p}^0_i) - (\mathbf{v}_j + s_j\mathbf{\dot p}^0_j),\\
\Delta \mathbf v^M_{ij} & = (\mathbf{v}_i + s_i\mathbf{\dot p}_i+V_{\|}\mathbf{p}_i) - (\mathbf{v}_j + s_j\mathbf{\dot p}_j+V_{\|}\mathbf{p}_j) \\
& = (\mathbf{v}_i + s_i\mathbf{\dot p}^0_i+V_{\|}\mathbf{p}^0_i + s_i V_{\|} \bm\kappa_i) - (\mathbf{v}_j + s_j\mathbf{\dot p}^0_j+V_{\|}\mathbf{p}^0_j+ s_j V_{\|} \bm\kappa_j).
\end{split}
\end{equation} 
Here $\mathbf{v}_i$ is the center-of-mass velocity of filament segment $i$, $s_i\mathbf{\dot p}^0_i$ is the rotational velocity of the crosslinker/motor head, $V_{\|}$ is the magnitude of motor head velocity, and $V_{\|}\mathbf{p}^0_i$ denotes the relative velocity with respect to the filament. 

Actin filaments continuously nucleate, assemble at their barbed  ($+$) ends, and disassemble at their pointed ($-$) ends. Due to the growth of filaments at the barbed end, fewer crosslinkers/motors bind to the barbed end compared to the pointed end (\cref{section:distribution}). We linearize the distribution of crosslinkers/motors on the filament segment and express $C^{X/M}(s_i, s_j)$ as
\begin{equation}
\label{eq:C}
\begin{split}
C^{X/M}(s_i, s_j) &= C^{X/M}_0 + C^{X/M}_1 (s_i + s_j ), \\
C^{X/M}_0 &= \frac{C^{X/M}_+ + C^{X/M}_{-}}{2},\\
C^{X}_1 &= \frac{C^{X}_+-C^{X}_{-}}{2L}, \\
C^M_1 &= 0,
\end{split}
\end{equation} 
where $C^{X/M}_+$ and $C^{X/M}_-$ are the crosslinker/motor concentrations at the segment ends close to the barbed and pointed ends. $C^X_1$ and $C^M_1$ account for deviations from uniform distributions along the filament. 
The choice of $C^M_1 = 0$ is based on the fact that the distribution of myosin motors is near uniform compared with that of crosslinkers in our simulations (\cref{section:distribution}).
We can express \cref{eq:stress_integration} using \cref{eq:force_velocity,eq:vij,eq:C}, 
\begin{equation}
\begin{split}
\label{eq:stress}
\mathbf\Sigma^{X}(\mathbf x) = &\frac{\gamma}{2} C^{X}_0 L^2 \sum_{ij} \int_{\Omega(x)}  \,d^3 \mathbf x^{\prime} \delta(\mathbf x_{i}-\mathbf x) \delta (\mathbf x_{j}-\mathbf x^{\prime})(\mathbf x_i - \mathbf x_j)(\mathbf{v}_i - \mathbf{v}_j ) \\
&+ \frac{\gamma}{2} C^{X}_0 \frac{L^4}{12} \sum_{ij} \int_{\Omega(x)}  \,d^3 \mathbf x^{\prime} \delta(\mathbf x_{i}-\mathbf x) \delta (\mathbf x_{j}-\mathbf x^{\prime})
(\mathbf p^0_i  \mathbf {\dot p}^0_i  + \mathbf p^0_j \mathbf {\dot p}^0_j)\\
&+ \frac{\gamma}{2} C^{X}_0 \frac{L^4}{12} \sum_{ij} \int_{\Omega(x)}  \,d^3 \mathbf x^{\prime} \delta(\mathbf x_{i}-\mathbf x) \delta (\mathbf x_{j}-\mathbf x^{\prime})
(\frac{\bm\kappa_i}{2} -\frac{\bm\kappa_j}{2})(\mathbf{v}_i - \mathbf{v}_j )  \\
& + \frac{\gamma}{2} C^{X}_1 \frac{L^4}{12} \sum_{ij} \int_{\Omega(x)}  \,d^3 \mathbf x^{\prime} \delta(\mathbf x_{i}-\mathbf x) \delta (\mathbf x_{j}-\mathbf x^{\prime})(\mathbf p^0_i - \mathbf p^0_j)(\mathbf{v}_i - \mathbf{v}_j ) ,\\
\mathbf\Sigma^{M}(\mathbf x) = &\frac{\gamma}{2} C^{M}_0 L^2 \sum_{ij} \int_{\Omega(x)}  \,d^3 \mathbf x^{\prime} \delta(\mathbf x_{i}-\mathbf x) \delta (\mathbf x_{j}-\mathbf x^{\prime})(\mathbf x_i - \mathbf x_j)(\mathbf{v}_i - \mathbf{v}_j +V_{\|}\mathbf{p}^0_i - V_{\|}\mathbf{p}^0_j) \\
&+ \frac{\gamma}{2} C^{M}_0 \frac{L^4}{12} \sum_{ij} \int_{\Omega(x)}  \,d^3 \mathbf x^{\prime} \delta(\mathbf x_{i}-\mathbf x) \delta (\mathbf x_{j}-\mathbf x^{\prime})
(\mathbf p^0_i  \mathbf {\dot p}^0_i  + \mathbf p^0_j \mathbf {\dot p}^0_j)\\
&+ \frac{\gamma}{2} C^{M}_0 \frac{L^4}{12} \sum_{ij} \int_{\Omega(x)}  \,d^3 \mathbf x^{\prime} \delta(\mathbf x_{i}-\mathbf x) \delta (\mathbf x_{j}-\mathbf x^{\prime})
[(\frac{\bm\kappa_i}{2} -\frac{\bm\kappa_j}{2})(\mathbf{v}_i - \mathbf{v}_j +V_{\|}\mathbf{p}^0_i - V_{\|}\mathbf{p}^0_j) +  V_{\|} (\bm\kappa_i \mathbf {p}^0_i - \bm\kappa_j \mathbf {p}^0_j)] \\
& + \frac{\gamma}{2} C^{M}_1 \frac{L^4}{12} \sum_{ij} \int_{\Omega(x)}  \,d^3 \mathbf x^{\prime} \delta(\mathbf x_{i}-\mathbf x) \delta (\mathbf x_{j}-\mathbf x^{\prime})(\mathbf p^0_i - \mathbf p^0_j)(\mathbf{v}_i - \mathbf{v}_j +V_{\|}\mathbf{p}^0_i - V_{\|}\mathbf{p}^0_j) .
\end{split}
\end{equation} 
Integrands proportional to either $s_{i/j}$ or $s_{i/j}^3$ do not contribute by symmetry. This expression is an extension of Eq.\ S49 in Ref.~\onlinecite{foster2022active} to account for the contribution of the local curvature $\bm\kappa_i$ to stress.  

To simplify these expressions, we follow Ref.~\onlinecite{foster2022active} to introduce the mass density for a filament segment with length $L$ as $\rho(\mathbf x) =  L \sum_i \delta(\mathbf x-\mathbf x_i)$, a rotational rate tensor $\mathcal{H}(\mathbf x) = \mathbf p^0_i \mathbf {\dot p}^0_i$, and a tensor that accounts for the coupling between filament orientation and velocity $\mathcal{J}(\mathbf x) = \mathbf p^0_i(\mathbf v_i - \mathbf{v(x)}) $. We also define tensors for the coupling between actin curvature and crosslinker/motor head velocity $\mathcal{K}^{X}(\mathbf x) = \frac{\bm \kappa_i}{2} (\mathbf v_i - \mathbf{v(x)}) $ and $\mathcal{K}^{M}(\mathbf x) = \frac{\bm \kappa_i}{2} (\mathbf v_i - \mathbf{v(x)}) +  \frac{3 \bm \kappa_i}{2} V_{\|} \mathbf{p}^0_i $. We compute the double sums in \cref{eq:stress} to obtain the stress tensors induced by crosslinkers and motors:
\begin{equation}
\label{eq:stress2}
\begin{split}
\mathbf\Sigma^X(\mathbf x) = &\frac{\gamma^X}{2} C^{X}_0 \int_{\Omega(x)}  \,d^3 \mathbf x^{\prime} (\mathbf x - \mathbf x^{\prime}) \rho(\mathbf x) \rho(\mathbf x^{\prime})( \mathbf v (\mathbf x) - \mathbf v (\mathbf x^{\prime}) ) \\
&+ \frac{\gamma^X}{2} C^{X}_0 \frac{L^2}{12} \int_{\Omega(x)}  \,d^3 \mathbf x^{\prime} \rho(\mathbf x) \rho(\mathbf x^{\prime})(\mathcal{H}  (\mathbf x) + \mathcal{H} (\mathbf x^{\prime}) ) \\
&+ \frac{\gamma^X}{2} C^{X}_0 \frac{L^2}{12} \int_{\Omega(x)}  \,d^3 \mathbf x^{\prime} \rho(\mathbf x) \rho(\mathbf x^{\prime})(\mathcal{K}^X  (\mathbf x) + \mathcal{K}^X (\mathbf x^{\prime}) ) \\
&+ \frac{\gamma^X}{2} C^{X}_1 \frac{L^2}{12} \int_{\Omega(x)}  \,d^3 \mathbf x^{\prime}  \rho(\mathbf x) \rho(\mathbf x^{\prime})
(\mathcal{J} (\mathbf x) + \mathcal{J} (\mathbf x^{\prime}) ),
\\
\mathbf\Sigma^M(\mathbf x) = &\frac{\gamma^M}{2}  C^{M}_0 \int_{\Omega(x)}  \,d^3 \mathbf x^{\prime} (\mathbf x - \mathbf x^{\prime}) \rho(\mathbf x) \rho(\mathbf x^{\prime})( \mathbf v (\mathbf x) - \mathbf v (\mathbf x^{\prime}) ) \\
&+ \frac{\gamma^M}{2} C^{M}_0 \frac{L^2}{12} \int_{\Omega(x)}  \,d^3 \mathbf x^{\prime} \rho(\mathbf x) \rho(\mathbf x^{\prime})(\mathcal{H}  (\mathbf x) + \mathcal{H} (\mathbf x^{\prime}) ) \\
&+ \frac{\gamma^M}{2} C^{M}_0 \frac{L^2}{12} \int_{\Omega(x)}  \,d^3 \mathbf x^{\prime} \rho(\mathbf x) \rho(\mathbf x^{\prime})(\mathcal{K}^M  (\mathbf x) + \mathcal{K}^M (\mathbf x^{\prime}) ) \\
&+ \frac{\gamma^M}{2} C^{M}_1 \frac{L^2}{12} \int_{\Omega(x)}  \,d^3 \mathbf x^{\prime}  \rho(\mathbf x) \rho(\mathbf x^{\prime})
(\mathcal{J} (\mathbf x) + \mathcal{J} (\mathbf x^{\prime}) + V_{\|}\frac{2}{3}\mathcal{I} ).
\end{split}
\end{equation} 
Further expanding $x^{\prime}$ around $x$, we can rewrite \cref{eq:stress2} as
\begin{equation}
\label{eq:stress3}
\begin{split}
\mathbf\Sigma^X(\mathbf x) = &\gamma^X  C^{X}_0 \int_{\Omega(x)}  \,d^3 \mathbf x^{\prime} (\mathbf x - \mathbf x^{\prime})^2 \rho(\mathbf x)^2 \div  \mathbf { v(x)} \\
&+ \gamma^X  C^{X}_0 \frac{L^2}{12} \int_{\Omega(x)}  \,d^3 \mathbf x^{\prime} \rho(\mathbf x)^2 \mathcal{H}  (\mathbf x) \\
&+ \gamma^X  C^{X}_0 \frac{L^2}{12} \int_{\Omega(x)}  \,d^3 \mathbf x^{\prime} \rho(\mathbf x)^2 
\mathcal{K}^X  (\mathbf x) \\
&+ \gamma^X  C^{X}_1 \frac{L^2}{12} \int_{\Omega(x)}  \,d^3 \mathbf x^{\prime} \rho(\mathbf x)^2 
\mathcal{J}  (\mathbf x), \\
\mathbf\Sigma^M(\mathbf x) =&\gamma^M  C^{M}_0 \int_{\Omega(x)}  \,d^3 \mathbf x^{\prime} (\mathbf x - \mathbf x^{\prime})^2 \rho(\mathbf x)^2 \div \mathbf { v(x)} \\
&+ \gamma^M  C^{M}_0 \frac{L^2}{12} \int_{\Omega(x)}  \,d^3 \mathbf x^{\prime} \rho(\mathbf x)^2 \mathcal{H}  (\mathbf x) \\
&+ \gamma^M C^{M}_0 \frac{L^2}{12} \int_{\Omega(x)}  \,d^3 \mathbf x^{\prime}  \rho(\mathbf x)^2 \mathcal{K}^M (\mathbf x)\\
&+ \gamma^M C^{M}_1 \frac{L^2}{12} \int_{\Omega(x)}  \,d^3 
\mathbf x^{\prime}  \rho(\mathbf x)^2 (\mathcal{J}(\mathbf x)+V_{\|}\frac{1}{3}\mathcal{I}).
\end{split}
\end{equation}
We follow the derivation in Ref.~\onlinecite{foster2022active} and derive the expression of $\mathcal{H}  (\mathbf x) = \frac{1}{3}\div \mathbf u - \frac{1}{9} Tr(\div \mathbf u)$ using the force and torque balance on a single filament. The first two terms in both $\mathbf\Sigma^X(\mathbf x)$ and $\mathbf\Sigma^M(\mathbf x)$ of \cref{eq:stress3} depend on the divergence of the velocity field $\div \mathbf u$ and are viscous-like terms. Thus, we arrive at viscosity $\eta \propto \rho ^2 (\gamma^M  C^{M}_0 +\gamma^X  C^{X}_0) $. 

The third and fourth terms in $\mathbf\Sigma^{X/M}(\mathbf x)$ of \cref{eq:stress3} are the active stress. To derive an expression for $\mathcal{J}  (\mathbf x)$, we follow Ref.~\onlinecite{foster2022active} and consider the force on a single filament. 
We can write the total force $F_{i}$, 
\begin{equation}
\label{eq:totalF_single}
\begin{split}
F_{i} &=\sum_{j} {\int_{-\frac{L}{2}}^{\frac{L}{2}} \,ds_i} 
{\int_{-\frac{L}{2}}^{\frac{L}{2}} \,ds_j}  
{\int_{\Omega(x_i)} \,d^3\mathbf x^{\prime}} \delta (\mathbf x_{j}-\mathbf x^{\prime})  (C^{M}_0 F^{M}_{ij} + C^{X}_0 F^{X}_{ij})\\
&= L^2 \sum_{j} {\int_{\Omega(x_i)} \,d^bythbf x^{\prime}}  \delta (\mathbf x_{j}-\mathbf x^{\prime})  (C^{M}_0 F^{M}_{ij} + C^{X}_0 F^{X}_{ij}).
\end{split}
\end{equation} 
Note that the force induced by the nonuniformly distributed crosslinkers involves the integration of $s_{i/j}$ terms, which do not contribute by symmetry. Using \cref{eq:force_velocity}, the total force can thus be expressed as,
\begin{equation}
\label{eq:Fi2}
\begin{split}
F_{i} = - L^2 \sum_{j}  \int_{\Omega(x_i)} \,&d^3\mathbf x^{\prime} \delta (\mathbf x_{j}-\mathbf x^{\prime})  \left[ ( C^{M}_0 \gamma^M +  C^{X}_0 \gamma^X )\times \right. \\
&\left.((\mathbf{v}_i + s_i\mathbf{\dot p}^0_i) - (\mathbf{v}_j + s_j\mathbf{\dot p}^0_j)) + C^{M}_0 \gamma^M V_{\|} (\mathbf{p}^0_i - \mathbf{p}^0_j +  s_i \bm\kappa_i - s_j \bm\kappa_j) \right].
\end{split}
\end{equation} 
Since the terms with $s_{i/j}\mathbf{\dot p}^0_{i/j}$ and $ s_{i/j} \bm\kappa_{i/j}$ also integrate to zero, we can further simplify \cref{eq:Fi2} to
\begin{equation}
\label{eq:Fi3}
F_{i} = - L^2 \sum_{j}  {\int_{\Omega(x_i)} \,d^3\mathbf x^{\prime}}  \delta (\mathbf x_{j}-\mathbf x^{\prime})  \left[ (C^{M}_0 \gamma^M + C^{X}_0 \gamma^X ) (\mathbf{v}_i -  \mathbf{v}_j) + C^{M}_0 \gamma^M V_{\|} (\mathbf{p}^0_i - \mathbf{p}^0_j) \right ].
\end{equation}
We can sum over $j$ in \cref{eq:Fi3} and obtain the following expression, 
\begin{equation}
\label{eq:Fi4}
F_{i} = - L  {\int_{\Omega(x_i)} \,d^3\mathbf x^{\prime}} \rho(\mathbf x^{\prime})
\left[ (C^{M}_0 \gamma^M + C^{X}_0 \gamma^X ) (\mathbf{v}_i -  \mathbf{v}(\mathbf x^{\prime})) + C^{M}_0 \gamma^M V_{\|} \mathbf{p}^0_i  \right ].
\end{equation}
Under steady flow conditions, the force on a single filament is balanced, so $F_i =0$, and we obtain
\begin{equation}
\label{eq:Fi5}
\mathbf v_i - \mathbf{v(x)}  =   -V_{\|} \mathbf p^0_i \frac{C^{M}_0 \gamma^M}{C^{M}_0 \gamma^M+ C^{X}_0 \gamma^X},
\end{equation}
and 
\begin{equation}
\label{eq:J}
\mathcal{J}(\mathbf x) = \mathbf p^0_i(\mathbf v_i - \mathbf{v(x)}) = -V_{\|} \mathbf p^0_i \mathbf p^0_i \frac{C^{M}_0 \gamma^M}{C^{M}_0 \gamma^M+ C^{X}_0 \gamma^X} = -\frac{1}{3} V_{\|} \frac{C^{M}_0 \gamma^M}{C^{M}_0 \gamma^M+ C^{X}_0 \gamma^X} \mathcal{I}.
\end{equation}
\cref{eq:Fi5,eq:J} indicate that the translational velocity of actin filaments is proportional to motor head velocity and is directed toward their pointed ends. The active stress term due to the non-uniform distribution of crosslinkers induced by actin turnover (the last term in $\mathbf\Sigma^X(\mathbf x)$  of \cref{eq:stress3}) is expressed as  $\gamma^X   C^{X}_1 \frac{L^2}{12} \int_{\Omega(x)}  \,d^3 \mathbf x^{\prime} \rho(\mathbf x)^2  \mathcal{J}  (\mathbf x)$. This term is the coupling between the translation of actin filaments and the accumulation of crosslinkers on the pointed ends of actins. 
Such terms can also contribute and lead to buckling. Consider for example, two filaments in an antiparallel configuration. 
The translational velocity of actin filaments, due to motor action, drives the sliding of two filaments passing each other. The accumulation of crosslinkers at the pointed however prevents such sliding leading to buckling.

\cref{eq:Fi5} also allows us to rewrite $\mathcal{K}^{M/X}$.
\begin{equation}
\begin{split}
\label{eq:kappa}
\mathcal{K}^{X}(\mathbf x) &= - \frac{V_{\|} \bm \kappa_i \mathbf p^0_i}{2}  \frac{C^{M}_0 \gamma^M}{C^{M}_0 \gamma^M+ C^{X}_0 \gamma^X}, \\
\mathcal{K}^{M}(\mathbf x) &= - \frac{V_{\|} \bm \kappa_i \mathbf p^0_i}{2}\frac{C^{M}_0 \gamma^M}{C^{M}_0 \gamma^M+ C^{X}_0 \gamma^X} +  \frac{3 V_{\|} \bm \kappa_i \mathbf p^0_i}{2}
\end{split}
\end{equation}
We now assume that $\bm \kappa_i p_i = -\frac{1}{3}\lambda \mathcal{I}$, where $\lambda$ is a constant that increases with curvature. Then, $\mathcal{K}^{M/X}$ can be expressed as, 
\begin{equation}
\begin{split}
\label{eq:final_kappa}
\mathcal{K}^{X}(\mathbf x) &= \frac{\lambda V_{\|} }{6}  \frac{C^{M}_0 \gamma^M}{C^{M}_0 \gamma^M+ C^{X}_0 \gamma^X} \mathcal{I}, \\
\mathcal{K}^{M}(\mathbf x) &= \frac{\lambda V_{\|} }{6}  \frac{C^{M}_0 \gamma^M}{C^{M}_0 \gamma^M+ C^{X}_0 \gamma^X} \mathcal{I} -  \frac{ \lambda V_{\|} }{2} \mathcal{I}.
\end{split}
\end{equation}
Substituting into the active stress terms in \cref{eq:stress3},
\begin{equation}
\begin{split}
\label{eq:sigma_a_final}
\mathbf{\Sigma_a}(\mathbf x) =& \frac{L^2}{36} \frac{4 \Pi R^3}{3}\rho(\mathbf x)^2 V_{\|}   \left[\frac{\lambda \gamma^X  C^X_0 }{2}  \frac{C^{M}_0 \gamma^M}{C^{M}_0 \gamma^M+ C^{X}_0 \gamma^X} \mathcal{I} 
 -  \gamma^X C^X_1 \frac{C^{M}_0 \gamma^M}{C^{M}_0\gamma^M+ C^{X}_0 \gamma^X} \mathcal{I}\right. \\
&\left. + \frac{\lambda \gamma^M  C^M_0 }{2}  \left(\frac{C^{M}_0 \gamma^M}{C^{M}_0 \gamma^M+ C^{X}_0 \gamma^X} - 3 \right) \mathcal{I} - \gamma^M C^M_1 \left(\frac{C^{M}_0 \gamma^M}{C^{M}_0\gamma^M+ C^{X}_0 \gamma^X} -1\right) \mathcal{I}\right].
\end{split}
\end{equation}
In our simulations, the concentrations of crosslinkers along filaments are at least 30 times those of motors, so $C_0^M \ll C_0^X$ and $C_1^M = 0$ (SI section 2F) and we can neglect the last two terms. We thus obtain the following expression for the active stress: 
\begin{equation}
\label{eq:sigma_a_final}
\Sigma_a \propto \rho ^2  \left(\frac{\lambda}{2}C_0^X \gamma^X- C_1^X \gamma^X\right) \left(V_{\|}\frac{ C_0^M \gamma^M}{C_0^M \gamma^M + C_0^X \gamma^X}\right).
\end{equation}
Combining this with $\eta \propto \rho ^2 (\gamma^M C_0^M +\gamma^X C_0^X) $, we obtain an expression for the contractility:
\begin{equation}
\label{eq:contractility_final}
d \propto  V_{\|} \frac{ (\frac{\lambda}{2}C_0^X \gamma^X- C_1^X \gamma^X)C_0^M \gamma^M}{(C_0^M \gamma^M + C_0^X \gamma^X)^2},
\end{equation}
which is Eq.\ 5 in the main text. \par
The concentration of myosin is $C_0^M = N^M/(N^A L)$, where $N^M$ is the number of myosins, $N^A$ is the number of actin segments, and $L$ is the length of each segment. 
Combining with the mass density of actin $\rho  = (N^A L)/V$, we express the myosin density as $\rho_M  = N^M/V  = C_0^M \rho$. 
We include a threshold $\rho_0$ and obtain that $C_0^M=\rho_M / (\rho-\rho_0)$. Inserting this expression into Eq.\ 5 results in Eq.\ 6 of the main text. \par
We define that $\bm \kappa_i p_i = -\frac{1}{3}\lambda \mathcal{I}$. If we denote the central angle of the curvature formed by an actin segment as $\theta$, we can derive that $\lambda/2 = 3(1-\cos\theta)/(2L)$, which takes on values within the range of [0,3/L] and remains positive. 
The curvature map intensity 0.27 (see Fig.\ 4a in the main text), which is scaled by the maximum curvature 4.79 $\mu$m$^{-1}$, corresponds to $\theta = 0.26$ and
$\lambda/2 \approx 1/(10 L)$. 

The ratio between non-uniform and uniform distributed crosslinkers $\frac{C_1^X}{C_0^X} = \frac{C^{X}_+-C^{X}_{-}}{L (C^{X/M}_+ + C^{X/M}_{-})}$, is in the range of [-1/($L_a$), 1/($L_a$)], where $L_a$ is the full length of the actin filament. We derive these two boundaries assuming zero concentration of crosslinkers at one end of the filament and 2$C_0^X$ concentration at the other ends. The full length of actin filaments falls within the range of 5 to 17 $\mu m$ (see Section S2F).
\subsection{Distribution of crosslinkers and motors along the filaments}
\label{section:distribution}
\begin{figure*}[hbp]
\centering
\includegraphics[width=0.65\textwidth,clip=true]{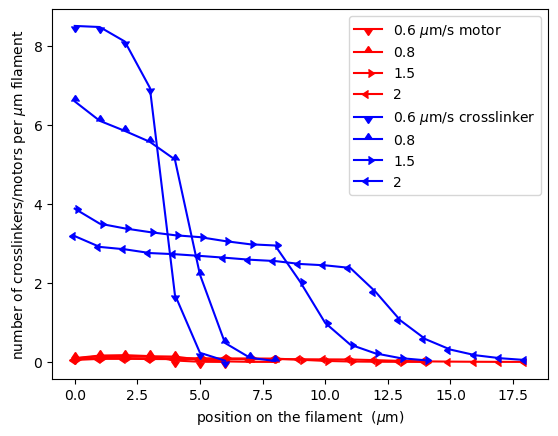}
\caption{Numbers of crosslinkers (blue) and motors (red) along the actin filaments. The filament lengths reached at 8.5 s are 5.1, 6.8, 12.75, and 17 $\mu$m for assembly/disassembly rates of 0.6, 0.8, 1.5, 2 $\mu$m/s. Each curve is generated from 40 frames.}
\label{fig:distribution}
\end{figure*}

We compute the number of crosslinkers/motors per unit length along filaments for various assembly/disassembly rates. \cref{fig:distribution} demonstrates that crosslinkers are nonuniformly distributed along the actin filament because of the shorter existing time of the actin segment near the barbed end. The motors are sparsely distributed and can be considered as evenly distributed along the filaments.
In the theory above, we approximate the nonuniform distribution by linearizing the distribution of crosslinkers from the barbed end to the pointed end. 
\clearpage

\clearpage
\bibliographystyle{unsrt}
\bibliography{references}